\documentclass[aps,floatfix,showpacs,prl,twocolumn,superscriptaddress,nofootinbib]{revtex4} 
\usepackage{graphicx} 
\usepackage{dcolumn} 
\usepackage{amsmath} 
 
\begin{document} 

\title{Origin of the anomalous long lifetime of $^{14}$C
}

\author{P. Maris}
\affiliation{Department of Physics and Astronomy,
Iowa State University, Ames, Ia 50011-3160, USA}
\author{J.P. Vary}
\affiliation{Department of Physics and Astronomy,
Iowa State University, Ames, Ia 50011-3160, USA}
\author{P. Navr\'atil}
\affiliation{TRIUMF, 4004 Wesbrook Mall, Vancouver BC, V6T 2A3, Canada}
\affiliation{Lawrence Livermore National Laboratory,
L-414, P.O. Box 808, Livermore, CA   94551, USA}
\author{W.\ E.\ Ormand}
\affiliation{Lawrence Livermore National Laboratory,
L-414, P.O. Box 808, Livermore, CA   94551, USA}
\affiliation{Department of Physics and Astronomy, 
Michigan State University, East Lansing, MI  48824, USA}
\author{H.\ Nam}
\affiliation{Oak Ridge National Laboratory,
P.O. Box 2008, Oak Ridge, TN  37831,  USA}
\author{D.\ J.\ Dean}
\affiliation{Oak Ridge National Laboratory,
P.O. Box 2008, Oak Ridge, TN  37831,  USA}

\begin{abstract}
We report the microscopic origins of the anomalously suppressed beta
decay of $^{14}$C to $^{14}$N using the 
{\it ab initio} no-core shell model (NCSM) with the Hamiltonian from
chiral effective field theory (EFT) including three-nucleon force (3NF) terms. 
The 3NF induces unexpectedly large 
cancellations within the $p$-shell between contributions to beta decay, which reduce 
the traditionally large contributions 
from the NN interactions by an order of magnitude, leading to the long lifetime of $^{14}$C.

\end{abstract}


\maketitle


The measured lifetime of $^{14}$C, $ 5730 \pm 30$ years, is a valuable
chronometer for many practical applications ranging from archeology 
to physiology.
It is anomalously long
compared to lifetimes of other light nuclei undergoing the same decay
process, allowed Gamow-Teller (GT) beta-decay, and it poses a major
challenge to theory since traditional realistic
nucleon-nucleon (NN) interactions alone appear insufficient to produce 
the effect\cite{aroua}.  Since the transition operator, in leading
approximation, depends on the nucleon spin and isospin
but not the spatial coordinate, this decay provides
a precision tool to inspect selected features of the initial
and final states. To convincingly explain this strongly inhibited transition,
we need a microscopic
description that introduces all physically-relevant 14-nucleon
configurations in the initial and final states and a realistic Hamiltonian
that governs the configuration mixing.

We report the first no-core solutions of $^{14}$C and
$^{14}$N using a Hamiltonian with firm ties to the
underlying theory of the strong interaction, Quantum Chromodynamics
(QCD), which allows us to isolate the key canceling contributions
involved in this beta decay.  We find that the three-nucleon force
(3NF) of chiral perturbation theory (ChPT) plays a major role in
producing a transition rate that is near zero, needed for the
anomalous long lifetime.  
A chiral 3NF with coupling constants consistent with other works 
and within their natural range can provide the precise lifetime.
This indicates that corrections to the lifetime
that arise from increasing the basis space, from including
additional many-body interactions and from corrections 
to the GT operator in ChPT  \cite{Vaintraub:2009mm} may
be absorbed into an allowed choice of the 3NF.

Our work features two major advances over recent alternative
explanations \cite{HKW,Brown-Rho}:
(1) we treat all nucleons on the same dynamical footing with the no-core
shell model (NCSM) \cite{NCSM}, and (2) we include the 3NF 
of ChPT \cite{Weinberg} as a full
3-nucleon interaction.
This follows previous work detailing the structure and electroweak properties of
selected A=10-13 nuclei \cite{nav} with the same chiral NN + 3NF. 
We also establish a foundation for future work on the GT
transitions to excited A=14 states \cite{negret}.

ChPT provides a theoretical framework for inter-nucleon interactions
based on the underlying symmetries of QCD.  Beginning with pionic or the nucleon-pion system
\cite{bernard95} one works consistently with systems of increasing
nucleon number \cite{ORK94,Bira,bedaque02a}.  One makes use of the
explicit and spontaneous breaking of chiral symmetry to 
expand the strong interaction in terms of a generic small momentum.
The ChPT expansion conveniently
divides the interactions into perturbative and non-perturbative
elements.  The latter are represented by a finite set of constants at
each order of perturbation theory that are not presently obtainable
from QCD, but can be fixed by experiment.  These constants should be
of order unity (``naturalness'').
Once determined, the resulting Hamiltonian predicts all other nuclear properties.
We have demonstrated that this reductive program works well for 
light nuclei when the 3NF is included.
\cite{nav}.

We adopt NN and 3NF potentials of ChPT 
\cite{Epelbaum,N3LO} and the no-core shell model 
(NCSM) to solve the many-body Schr\"odinger 
equation for A=14 nuclei while preserving all Hamiltonian 
symmetries \cite{NCSM, NavratilFB07,v3eff}.
The non-perturbative coupling constants of the 3NF, not
fixed by $\pi-N$ or NN data, are  $c_D$ for
the $N-\pi-NN$ contact term and $c_E$ for the $NNN$
contact term.  We previously fit $c_D$ and $c_E$
to A=3 binding energies and selected
$s$-shell and $p$-shell nuclear properties and showed the resulting
interactions worked well for A=10-13 nuclei \cite{nav}.  
Here, we present results for $(c_D,c_E)=(-0.2,-0.205)$ and for
$(-2.0,-0.501)$.
Both sets, labeled by their $c_D$ value below, are allowed 
by the naturalness criterium and fit the A=3 binding energies.
The former  also produces a precise fit to the triton half life
\cite{Gazit:2008ma}.  The latter produces the triton half life 
within $20 \%$ of experiment but is preferred by the $^{14}$C half life
as we show here.

We select a basis space of harmonic oscillator (HO)
single-particle states specified by the HO energy $\hbar\Omega$ and by
$N_{max}$,  the limit on the number of HO quanta, $\sum_i (2n_i
+ l_i)$, above the minimum for that nucleus.  $n_i$ is the
principal quantum number of the orbital of the $i^{th}$ nucleon and
$l_i$ is its orbital angular momentum. Thus $N_{max}=0$ is the smallest
basis space for each nucleus consistent with the Pauli principle. For
each choice of $N_{max}$ and $\hbar\Omega$ we carry out 
a well-established finite-basis
renormalization \cite{LSO} of the Hamiltonian to arrive at an
Hermitian effective 3-body Hamiltonian \cite{NCSM,v3eff}.  
All symmetries of the underlying
Hamiltonian are preserved throughout.  
We adopt $\hbar\Omega=14$ MeV,
which corresponds to the minimum in the binding energy 
obtained for $N_{max} =8$.
We have checked that results for $13 \leq \hbar\Omega \leq 18$ 
MeV do not qualitatively alter our conclusions.

Basis space dimensions grow rapidly with $N_{max}$ and provide a major
technical challenge \cite{MFDn}.
The many-body matrix dimension in the M-scheme basis
(total magnetic projection quantum number is fixed) for $^{14}$C
($^{14}$N) is 872,999,912 (1,090,393,922) at $N_{max}=8$ and the associated number of
non-vanishing 3NF matrix elements on and below the diagonal $\simeq 2.9
\times 10^{13}$ ($\simeq 3.9 \times 10^{13}$).

We obtained our results on the Jaguar supercomputer at
Oak Ridge National Lab \cite{Jaguar} 
using up to 35,778 hex-core processors (214,668 cores) 
and up to 6 hours of elapsed time for each set of low-lying eigenvalues and
eigenvectors.  The number of non-vanishing matrix elements exceeded
the total memory available and required matrix element recomputation
``on-the-fly'' for the iterative diagonalization process (Lanzos algorithm).

\begin{figure}[hbt]
\begin{center}
\includegraphics[width=0.95\columnwidth]{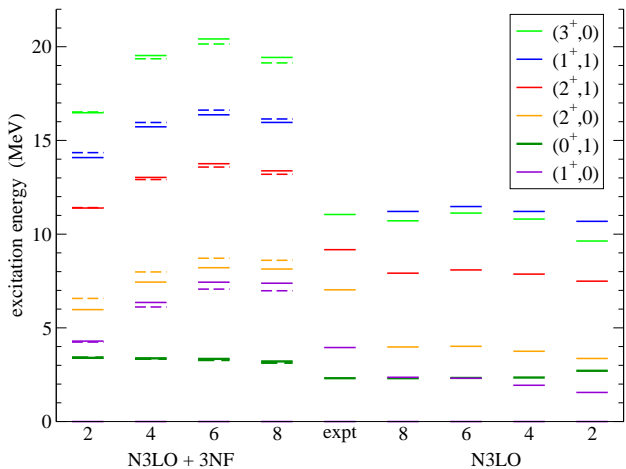}
\caption{(Color online) The excitation spectrum of $^{14}$N obtained in the NCSM
  using chiral interactions as a function of basis space cutoff
  $N_{max}$, the number that labels the columns. ($J^{\pi},T$) values are in the legend.
  The center column
  displays the experimental spectrum for $^{14}$N.  Columns to the
  right are obtained with the chiral NN interaction alone while those
  on the left are obtained with the chiral NN + 3NF using $c_D=-0.2$
  (solid lines) and $c_D=-2.0$ (dashed lines). \label{n14spectrum} }
\end{center}
\end{figure}
Fig.~\ref{n14spectrum} compares our spectra for $^{14}$N with experiment.
The $T=1$ isobaric analog states serve as good proxies for the $^{14}$C spectra.
We present the spectra as a function of the $N_{max}$ value (column labels) 
for both the chiral NN interaction alone 
(``N3LO'' columns) and the chiral NN + 3NF (``N3LO+3NF'' columns).  

The trends in excitation energies with 
increasing $N_{max}$ suggest reasonable convergence.  
We note that the level orderings are the same with and without the 3NF
in the largest basis spaces ($N_{max}=8$).
Overall, the spectra are more compressed with NN alone than with NN +
3NF, a pattern seen before \cite{nav,v3eff}.  
However, the 3NF appears to have over compensated for the compression
in the NN spectra when compared with experiment. 
On the other hand the 3NF improves the binding energies. 
The experimental ($^{14}$C, $^{14}$N)
binding energies are   
(105.28, 104.66) MeV.
Without the 3NF, we obtain (97.20, 96.36) MeV. 
When we include the 3NF the binding energies increase to (108.43, 108.47) MeV
for $c_D = -0.2$ and (108.30, 108.41) MeV for $c_D=-2.0$

We note that the binding energies for both $^{14}$C and $^{14}$N are
about 8 MeV or 8\% larger at $N_{max}=6$ than they are at $N_{max}=8$.
Similar calculations for lighter nuclei show that the convergence of
the binding energy is not monotonic, hampering an extrapolation to the
complete (infinite-dimensional) basis.  
On the other hand, the excitation spectra appear
rather stable upon reaching $N_{max}=8$.  We take
results in the largest available basis space as estimates of the
converged results.  
We also show in Fig.  \ref{n14spectrum} that the spectra are relatively insensitive 
to a significant range of values for ($c_D, c_E$).

Using Fermi's Golden rule for the transition rate, the
half life $T_{1/2}$ for $^{14}$C is given by
\begin{eqnarray}
  T_{1/2} &=& \frac{1}{f(Z,E_0)}\frac{2\pi^3 \hbar^7 {\rm ln} 2}{m_e^5 c^4 G_V^2}\frac{1}{{g_A}^2|M_{\rm GT}|^2}\, ,
  \label{EqHalfT}
\end{eqnarray}
where $M_{\rm GT}$ is the reduced GT matrix element; 
$f(Z,E_0)$ is the Fermi phase-space integral;
$E_0=156$ keV is the $\beta$ endpoint;
$G_V= 1.136 ~ 10^{-11}$ MeV$^{-2}$ is the weak vector coupling constant; 
$g_A=1.27$ is the axial vector coupling constant;
and $m_e$ is the electron mass.
$M_{\rm GT}$ for the
transition from the initial $^{14}C$ $(J^\pi,T) = (0^+, 1)$ ground
state ($\Psi_i$) to the $^{14}N$ $(1^+, 0)$ ground state ($\Psi_f$) is defined by the
spin-isopin operator $\sigma(k)\tau_+(k)$ acting on all nucleons, $k$:
\begin{eqnarray}
M_{\rm GT} &=& \sum_{k} \left<\Psi_f||\sigma(k)\tau_+(k)||\Psi_i\right>.
\label{gte}
\end{eqnarray}
Since the initial state has total spin zero, $M_{\rm GT}$ is equal to the
M-dependent GT matrix element $M_{\rm GT}^{M}$, 
\begin{eqnarray}
M_{\rm GT}^{M} 
&=& \sum_{\alpha,\beta}  \left<\alpha|\sigma\tau_+|\beta\right>  \rho_{\alpha\beta},
\label{gte2}
\end{eqnarray}
where $\left<\alpha|\sigma\tau_+|\beta\right>$ is the one-body matrix
element between HO single-particle states $\alpha$ and $\beta$, 
which is non-vanishing only when both single-particle states are in the same shell, 
and the one-body density matrix 
$ \rho_{\alpha\beta} \equiv \left<\Psi_f|a_{\alpha}^{\dagger}a_{\beta}|\Psi_i\right>$.

In order to reproduce the measured half life of $T_{1/2} \simeq 5730$
years, the GT matrix element must be anomalously small,
$|M_{\rm GT}^{M}| \simeq 2 \times 10^{-3}$, in contrast with a conventional
strong GT transition in a light nucleus 
with $|M_{\rm GT}| \simeq 1$.

\begin{figure}[hbt]
\begin{center}
\includegraphics[width=0.95\columnwidth]{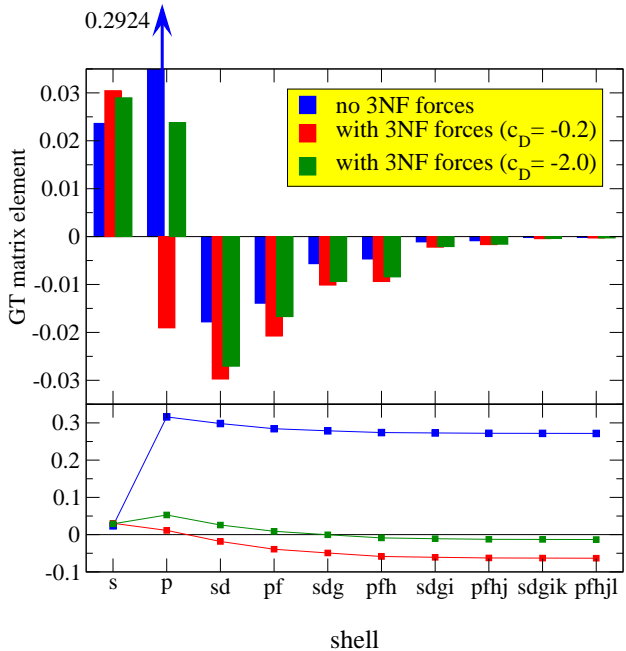}
\caption{(Color online) Contributions to the $^{14}$C beta decay matrix element as a
  function of the HO shell in which they are evaluated when the
  nuclear structure is described by the chiral interaction. Top panel
  displays the contributions with (two right bars of each triplet) and without
  (leftmost bar of each triplet) the 3NF at $N_{max}=8$.  All contributions
  are summed within the shell to yield a total for that shell.  The
  bottom panel displays the running sum of the GT contributions over
  the shells included in the sum. Note the order-of-magnitude
  suppression of the $p$-shell contributions arising from the 3NF.
\label{c14_beta}}
\end{center}
\end{figure}

In Fig.~\ref{c14_beta}, we present our main results for $M_{\rm GT}$. 
We decompose $M_{\rm GT}$ at $N_{max}=8$ into the contributions 
arising from each HO shell for two cases 
with the 3NF ($c_D = -0.2, -2.0$) and one without.
The largest effect occurs in the $p$-shell where, for both values
of $c_D$, the 3NF reduces
the contributions by an order-of-magnitude from the result with the
NN interaction alone.  The contributions of each of the 9
additional shells is enhanced by up to a factor of 2 by the 3NF.  The
cumulative contributions to the GT matrix element is displayed in the
lower panel of Fig. \ref{c14_beta} where one sees clearly the net
suppression to $M_{\rm GT}$ due to the 3NF.  

Looking into the detailed changes within the
$p$-shell, one finds that the 3NF introduces a systematic shift of
strength away from 1-body density matrix terms involving transitions
between the 0p$_{3/2}$ and the 0p$_{1/2}$ orbits to
1-body density matrix elements involving transitions within the
0p$_{1/2}$ orbits.  The shift is about 30\% of the magnitudes
of these 1-body density matrix terms and supports a recurring theme of
3NF's in $p$-shell nuclei - they play a significant role in the
spin-sensitive properties of spin-orbit pairs.  We note that our
observed shifts within the $p$-shell due to the 3NF is similar to what
Ref. \cite{HKW} accomplished with a density-dependent effective NN
interaction obtained by modeling leading contributions from the chiral
3NF.  However, our net contributions from other shells, which are
absent in Ref. \cite{HKW}, overwhelm the net $p$-shell contribution as
seen in Fig. \ref{c14_beta}.  That is, while the $s$-shell and
$sd$-shell contributions nearly cancel, all the shells above the
$sd$-shell contribute about a factor 2 greater than the
now-suppressed $p$-shell contribution.

To further understand the role of the 3NF, we can examine the
contributions to $M_{GT}$ in the LS-scheme where the single-particle
quantum numbers now involve the orbital angular momentum projection
$m_l$ and spin projection $m_s$ replacing the total angular momentum
projection $m_j$. This is a convenient representation for this
transition since $m_l$ and $m_s$ must be the same for incoming and
outgoing single-particle states.  For the $p$-shell contributions, the
resulting decomposition to LS-scheme yields results shown in Table
\ref{LS-table}. Note that there is nearly perfect cancellation between
the $m_l=0$ and $m_l=\pm1$ terms once the 3NF is included.

Given the overall effects on $M_{\rm GT}$ by inclusion of the 3NF as
seen through Fig. \ref{c14_beta} and in Table \ref{LS-table}, one may 
fine tune $c_D$ to reproduce the value $2 \times 10^{-3}$
consistent with the $^{14}$C lifetime.  This is analogous to the fine
tuning  of $c_D$ in Ref. \cite{Gazit:2008ma} to fit the $^3$H lifetime.  
Our estimate that the desired suppression
of $M_{\rm GT}$ occurs with $c_D \simeq -2.0$ illustrates this point.
Note that the spectra of $^{14}$C and $^{14}$N are rather
insensitive to this range of $c_D$ values.  We show the resulting
$M_{\rm GT}$ in the final row of Table \ref{LS-table}.

\begin{table}[htb]
\begin{center}
\begin{tabular}{|c||c|c|c|} \hline
($m_l$,$m_s$) & NN only & NN + 3NF & NN + 3NF\\ 
                            &                             & $c_D=-0.2$         & $c_D=-2.0$    \\ \hline
($1,+\frac{1}{2}$)  & $0.015$   & $0.009$         & $0.009$     \\ \hline
($1,-\frac{1}{2}$)   & $-0.176$  & $-0.296$       &$-0.280$ \\ \hline
($0,+\frac{1}{2})$  & $0.307$   & $0.277$        &$0.283$ \\ \hline
($0,-\frac{1}{2})$   & $0.307$   & $0.277$        &$0.283$ \\ \hline
($-1,+\frac{1}{2})$ & $-0.176$  & $-0.296$      &$-0.280$ \\ \hline
($-1,-\frac{1}{2})$  & $0.015$   & $0.009$        &$0.009$ \\ \hline
Subtotal                      &    $0.292$& $ -0.019$  &  $0.024$ \\ \hline
Total Sum                  &    $0.275$    & $-0.063$        &    $-0.013$ \\ \hline


\end{tabular}
\caption{Decomposition of $p$-shell contributions to $M_{GT}$ in the LS-scheme
for the beta decay of $^{14}$C without and with 3NF.  The 3NF is included at
two values of $c_D$ where $c_D\simeq-0.2$ is preferred by the $^3$H  lifetime
and $c_D\simeq-2.0$ is preferred by the $^{14}$C lifetime. The calculations
are performed in the $N_{max}=8$ basis space with $\hbar\Omega=14$ MeV.
\label{LS-table}}
\end{center}
\end{table}

Next, we studied the sensitivity at $(c_D, c_E)=(-2.0,-0.501)$
and at $N_{max}=6$ by successively setting each
to zero  while keeping the other fixed.
The larger effect on $M_{GT}$ appears with 
$c_D=0$.  However, the resulting shifts in the magnitude of $M_{GT}$ are approximately
proportional to the magnitude of the changes in  $c_D$ and $c_E$
which implies $M_{GT}$ has about the same sensitivity to each.
This sensitivity differs from that of Ref.~\cite{HKW} where 
the $c_E$ term was found to play a leading role in $M_{GT}$.

Since both appear natural, we conclude that the different values of $c_D$ 
from the beta decays of $^3$H and $^{14}$C accommodate the sum of
higher-order effects:
meson-exchange currents, additional ChPT interaction terms, 
and contributions from larger basis spaces. 
Additional 3NF terms, e.g.  those arising at the N$^3$LO
level of ChPT that do not impact the $^3$H lifetime \cite{Gardestig:2006hj}, 
and/or four
nucleon interaction terms may impact $^{14}$C and $^{14}$N.  

In order to further test the {\it ab initio} wavefunctions and the effects of the 3NF,
we present in Table \ref{M1-table} the contributions to selected
electromagnetic properties of $^{14}$N.  We see that the 3NF 
has substantial influence on the magnetic-dipole transition in
$^{14}$N from the ground state (GS) to the $(0^+,1)$ analog state of the $^{14}$C GS
though the mechanism for this suppression is quite different from the GT
transition.  In this M1 transition, we find
significant cancellation between the nucleon
spin and proton angular momentum contributions.  Without the 3NF, 
spin and angular momentum contributions add constructively.
Electric properties such as the charge radius ($RMS$) 
and the electric quadrupole moment ($Q$) 
are reduced 
when we include 3NF due, in part, to the increased binding.  Also, the GS
magnetic moment ($\mu$) is modified by 10\% when we include 3NF. 
We expect $RMS$ and $Q$ to increase with basis space size while 
the GS energy, $\mu$ and the B(M1) will be less effected.
The role of the 3NF on the electroweak properties of these nuclei 
is indeed multi-faceted. 

\begin{table}[htb]
\begin{tabular}{|c||c||c|c|c|} \hline
Observable  & Experiment     & NN only   & NN + 3NF  & NN + 3NF  \\ 
                     &    \cite{Expt}       &                 & $c_D=-0.2$ & $c_D=-2.0$        \\ \hline
 $RMS$      & $2.42(1)$    & $2.28$    &  $2.25$          &  $2.24$                 \\ \hline
  $Q$           &  $1.93(8)$   & $1.87$    &  $1.03$          &  $1.19$                \\ \hline
  $\mu$       & $0.404$       & $0.379$  &  $0.347$       &  $0.347$              \\ \hline
B(M1)       & $0.047(2)$     & $1.002$    &   $0.037$      &  $0.098$            \\ \hline
\end{tabular}
\caption{Properties of $^{14}$N without and with 3NF 
in the $N_{max}=8$ basis space with $\hbar\Omega=14$ MeV. 
The magnetic moments, which tend to converge 
rapidly, are obtained at $N_{max}=6$.  The point proton
root-mean-square radius (RMS) is quoted in fm.  
We corrected the measured charge radius (2.56(1) fm) 
for the finite proton charge contribution. 
The magnetic moment $\mu$ is in nuclear magnetons $e\hbar/2mc$; 
and the quadrupole moment is in $e^2\,fm^4$ (all for the GS).  
The B(M1) is the transition from the GS to the $(0^+,1)$ state (the
isobaric analog of the $^{14}$C GS).} 
\label{M1-table} 
\end{table}

In conclusion, the chiral 3NF in {\it ab initio} nuclear physics
produces a large amount of cancellation in the matrix element
$M_{\rm GT}$ governing the beta decay of $^{14}C$. This cancellation
signals a major signature of 3NF effects in the spin-isopin content of
the $0p$ orbitals in $^{14}$C and $^{14}$N.  The 3NF, particularly
through its longest range two-pion exchange component, modifies the
$0p$ spin-orbit pairs in the ground states of these nuclei to produce
the suppression of $M_{\rm GT}$ consistent with the anomalous
long lifetime of $^{14}$C.

This work was supported in part by US DOE Grants DE-FC02-09ER41582 (UNEDF SciDAC Collaboration), 
DE-FG02-87ER40371, and by US DOE Contract No. DE-AC52-07NA27344 and
No. DE-AC05-00OR22725 .  Computational resources were provided 
by Livermore Computing at LLNL and
by  a ``Petascale Early Science Award'' and an INCITE Award 
on the Jaguar supercomputer at the Oak Ridge Leadership Computing Facility at ORNL 
\cite{Jaguar}
which is supported by the DOE Office of Science under Contract 
DE-AC05-00OR22725.

\end{document}